\documentclass{IEEEtran}

\usepackage[T1]{fontenc}
\usepackage{fst}
\ifCLASSINFOpdf
\else
\fi
\usepackage{amsmath}
\usepackage{comment}
\usepackage[cmintegrals]{newtxmath}

\begin{document}
\title{Low-Resolution Precoding for Multi-Antenna Downlink Channels and OFDM}
\author{%
Andrei Nedelcu,
Fabian Steiner, and
Gerhard Kramer
\thanks{Date of current version \today.
This work was supported by the German Research Foundation (DFG) under Grant KR 3517/9-1. The results of this paper have been presented in part at the Workshop on Smart Antennas (WSA) 2018.}
\thanks{
Andrei Nedelcu  was with the Institute for Communications Engineering, Technical University of Munich (TUM), 80333 Munich, Germany. He is now with the Huawei Munich Research Center, 80992 Munich, Germany (e-mail: andrei.nedelcu2@huawei.com).

Fabian Steiner was with the Institute for Communications Engineering, Technical University of Munich (TUM), 80333 Munich, Germany. 

Gerhard Kramer is with the Institute for Communications Engineering, Technical University of Munich (TUM), 80333 Munich, Germany (e-mail: gerhard.kramer@tum.de).
}
}

\maketitle

\begin{abstract}
Downlink precoding is considered for multi-path multi-input single-output channels where the base station uses orthogonal frequency-division multiplexing and low-resolution signaling. A quantized coordinate minimization (QCM) algorithm is proposed and its performance is compared to other precoding algorithms including squared infinity-norm relaxation (SQUID), multi-antenna greedy iterative quantization (MAGIQ), and maximum safety margin precoding. MAGIQ and QCM achieve the highest information rates and QCM has the lowest complexity measured in the number of multiplications. The information rates are computed for pilot-aided channel estimation and
data-aided channel estimation. Bit error rates for a 5G low-density parity-check code confirm the information-theoretic calculations. Simulations with imperfect channel knowledge at the transmitter show that the performance of QCM and SQUID degrades in a similar fashion as zero-forcing precoding with high resolution quantizers.
\end{abstract}
 

\IEEEpeerreviewmaketitle

\section{Introduction}

\IEEEPARstart{M}{assive} multiple-input multiple-output (MIMO) base stations can serve many \acp{UE} with high spectral efficiency and simplified signal processing~\cite{marzetta_noncooperative_2010,ngo_energy_2013}. However, their implementation is challenging due to the cost and energy consumption of analog-to-digital and digital-to-analog converters (ADCs/DACs) and linear \acp{PA}. There are several approaches to lower cost. One approach is hybrid beamforming with analog beamformers in the \ac{RF} chain of each antenna and where the digital baseband processing is shared among \ac{RF} chains. Second, constant envelope waveforms permit using non-linear \acp{PA}. Third, all-digital approaches use low-resolution ADCs/DACs or low-resolution digitally controlled \ac{RF} chains. The focus of this paper is on the all-digital approach.

\subsection{Single-Carrier Transmission}
We study the multi-antenna downlink and \acp{UE} with one antenna each, a model referred to as \ac{MU-MISO}. Most works on low-cost precoding for \ac{MU-MISO} consider \ac{PSK} to lower the requirements on the \acp{PA}.
For instance, the early papers~\cite{Mohammed-Larsson-COMM13,Mohammed-Larsson-COMML13}(see also~\cite{Mohammed-Larsson-WCOMM12}) use iterative coordinate-wise optimization to choose transmit symbols from a \emph{continuous} \ac{PSK} alphabet for flat and frequency-selective (or multipath) fading, respectively. We remark that these papers do not include an optimization parameter (called $\alpha$ below, see~\eqref{eq:cost-function}) in their cost function that plays an important role at high \ac{SNR}, see~\cite{Joh05,bjornson_precoding_2014}. This parameter is related to linear \ac{MMSE} precoding.

Most works consider \emph{discrete} alphabet signaling. Perhaps the simplest approach, called \ac{QLP}, applies a linear precoder followed by one low-resolution quantizer per antenna\cite{Mezghani_2009,Mezghani_2009_G,Usman_2016,Saxena_2016_2,Kakkavas_2016,li_2017,Swindlehurst_2018,amodh_2020}. Our focus is on \ac{ZF}, and we use the acronyms LP-ZF and QLP-ZF, respectively, for unquantized \ac{ZF} and the \ac{QLP} version of \ac{ZF}.

More sophisticated approaches use optimization tools as in~\cite{Mohammed-Larsson-COMML13,Mohammed-Larsson-COMM13}. For example, the papers~\cite{Jedda_2016,jacobsson_quantized_2017,Wang-etal-WCOM18} use convex relaxation methods; \cite{Markus17,Shao_2018,Ang_Li_2018,Ang_Li_2021,Ang_Li_F_Liu_2021,tsinos2018,domouchtsidis2019} apply coordinate-wise optimization; \cite{Jedda-etal-17,Jedda_2018,Jedda_2017_2} develops a symbol-wise Maximum Safety Margin (MSM) precoder; \cite{Landau-deLamare17,Jacobsson_2018,Li-Fan-Masouros-IEEE-TrWire,Lopes-Landau-WSA20} use a branch-and-bound (BB) algorithm;
\cite{Shao_2019} uses a majorization-minimization algorithm; \cite{Sedaghat_2018} uses integer programming; and
\cite{Balatsoukas-etal-SPAWC19,Sohrabi_2020} use neural networks (NNs).
These references are collected in Table~\ref{table:references} together with the papers listed below on \ac{OFDM}. As the table shows, most papers focus on single-carrier and flat fading channels.

\begin{table*}[t]
	\center
	\setlength{\tabcolsep}{7pt} 
	\renewcommand{\arraystretch}{1.1} 
	\begin{tabular}{ |l|l||l|l|l|l| } 
		\cline{3-6}
		\multicolumn{2}{c|}{}
		& \multicolumn{4}{c|}{Precoding Algorithm}
		\\ \cline{3-6}
		\multicolumn{2}{c|}{}
		& \ac{QLP} & Convex & Coord.-Wise & Other (MSM,
		\\ \cline{1-2}
		\multicolumn{1}{|c|}{Modulation} &
		\multicolumn{1}{c||}{Fading} & 
		& Relaxation & Optimization & BB, NN, etc.)
		\\ \hline \hline
		\multirow{2}{*}{1 Carrier}
		& Flat
		& \cite{Mezghani_2009,Mezghani_2009_G,Usman_2016,Saxena_2016_2,Kakkavas_2016,li_2017,Swindlehurst_2018,amodh_2020}
		& \cite{Jedda_2016,jacobsson_quantized_2017,Wang-etal-WCOM18}
		& \cite{Markus17,Shao_2018,Ang_Li_2018,Ang_Li_2021,Ang_Li_F_Liu_2021,tsinos2018,domouchtsidis2019}
		& \cite{Jedda-etal-17,Jedda_2018,Landau-deLamare17,Jacobsson_2018,Li-Fan-Masouros-IEEE-TrWire,Lopes-Landau-WSA20,Shao_2019,Sedaghat_2018,Balatsoukas-etal-SPAWC19,Sohrabi_2020}
		\\ & Freq.-Sel.
		&
		& 
		&
		& \cite{Jedda_2017_2}
		\\ \hline
		OFDM & Freq.-Sel.
		& \cite{Jacobsson-etal-WCOM19}
		& \cite{Jacobsson-etal-ICT18}
		& \cite{Nedelcu-etal-WSA18,Tsinos-etal-Access20,Tsinos-etal-WC21}
		& \cite{Hela_MSM_OFDM,Mezghani_2022}
		\\ \hline
	\end{tabular}
	\caption{References for quantized precoding.}
	\label{table:references}
	\end {table*}
\subsection{Discrete Signaling and OFDM}
Our main interest is discrete-alphabet precoding for multipath channels with \ac{OFDM} as in 5G wireless systems. Precoding for \ac{OFDM} is challenging because the alphabet constraint is in the time domain after the inverse discrete Fourier transform (IDFT) rather than in the frequency domain. We further focus on using information theory to derive achievable rates. For this purpose, we consider two types of channel estimation at the receivers: pilot-aided channel estimation via \ac{PAT} and data-aided channel estimation.

Discrete-alphabet precoding for OFDM was treated in~\cite{Jacobsson-etal-WCOM19} that uses \ac{QLP} and low resolution DACs. A more sophisticated approach appeared in~\cite{Jacobsson-etal-ICT18} that applies a squared-infinity norm Douglas-Rachford splitting (SQUID) algorithm to minimize a quadratic cost function in the \emph{frequency} domain. The performance was illustrated via \ac{BER} simulations with convolutional codes and QPSK or 16-\ac{QAM} by using 1-3 bits of phase quantization.

The paper~\cite{Nedelcu-etal-WSA18} instead proposed an algorithm called multi-antenna greedy iterative quantization (MAGIQ) that builds on \cite{Markus17} and uses coordinate-wise optimization of a quadratic cost function in the \emph{time} domain. MAGIQ may thus be considered an extended version of~\cite{Mohammed-Larsson-COMML13} for \ac{OFDM} and discrete alphabets. Simulations showed that MAGIQ outperforms SQUID in terms of complexity and achievable rates. Another coordinate-wise optimization algorithm appeared in~\cite{Tsinos-etal-Access20,Tsinos-etal-WC21} that builds on the papers~\cite{tsinos2018,domouchtsidis2019}. The algorithm
is called \ac{CESLP} and it is similar to the refinement of MAGIQ presented here. The main difference is that, as in~\cite{Jacobsson-etal-ICT18}, the optimization in~\cite{Tsinos-etal-Access20,Tsinos-etal-WC21} uses a cost function in the frequency domain rather than the time domain. We remark that processing in the time domain has advantages that are described in Sec.~\ref{sec:algoFS}.

The MSM algorithm was extended to OFDM in~\cite{Hela_MSM_OFDM}. MSM works well at low and intermediate rates but MAGIQ outperforms MSM at high rates both in terms of complexity and achievable rates.
Finally, the recent paper~\cite{Mezghani_2022} uses generalized approximate message passing (GAMP) for OFDM.

\subsection{Contributions and Organization}
\label{subsec:contributions}
The contributions of this paper are as follows.
\begin{itemize}
	\item The analysis of MAGIQ in the workshop paper~\cite{Nedelcu-etal-WSA18} is extended to larger systems and more realistic channel conditions.
	\item Replacing the greedy antenna selection rule of MAGIQ with a fixed (round-robin) schedule is shown to cause negligible rate loss. The new algorithm is named \ac{QCM}.
	\item The performance of QLP-ZF, SQUID, MSM, MAGIQ, and \ac{QCM} are compared in terms of complexity (number of multiplications and iterations) and achievable rates.
	\item We develop an auxiliary channel model to compute achievable rates for pilot-aided and data-aided channel estimation. The models let one compare modulations, precoders, channels, and receivers.
	\item Simulations with a 5G NR \ac{LDPC} code~\cite{bae_abotabl_lin_song_lee_2019} show that the computed rate and power gains accurately predict the gains of standard channel codes.
	\item Simulations with imperfect channel knowledge at the base station show that the achievable rates of SQUID and \ac{QCM} degrade as gracefully as those of LP-ZF.
\end{itemize}

We remark that our focus is on algorithms that approximate ZF based on channel \emph{inversion}, i.e., there is no attempt to optimize transmit powers across subcarriers. This approach simplifies OFDM channel estimation at the receivers because the precoder makes all subcarriers have approximately the same channel magnitude and phase. For instance, a rapid and accurate channel estimate is obtained for each OFDM symbol by averaging the channel estimates of the subcarriers, see Sec.~\ref{sec:rate_computation}. Of course, it is interesting to develop algorithms for other precoders and for subcarrier power allocation.

This paper is organized as follows. Sec.~\ref{chap:system_model} introduces the baseband model and OFDM signaling. Sec.~\ref{sec:freq_sel_chan} describes the MAGIQ and \ac{QCM} precoders. Sec.~\ref{sec:rcs} develops theory for achievable rates, presents complexity comparisons, and reviews a model for imperfect \ac{CSI}. Sec.~\ref{sec:numerical} compares achievable rates and \acp{BER} with 5G NR LDPC codes. Sec.~\ref{sec:conclusion} concludes the paper.

\section{System Model}
\label{chap:system_model}
\begin{figure*}[t!]
	\centerline{\includegraphics[scale=1.1]{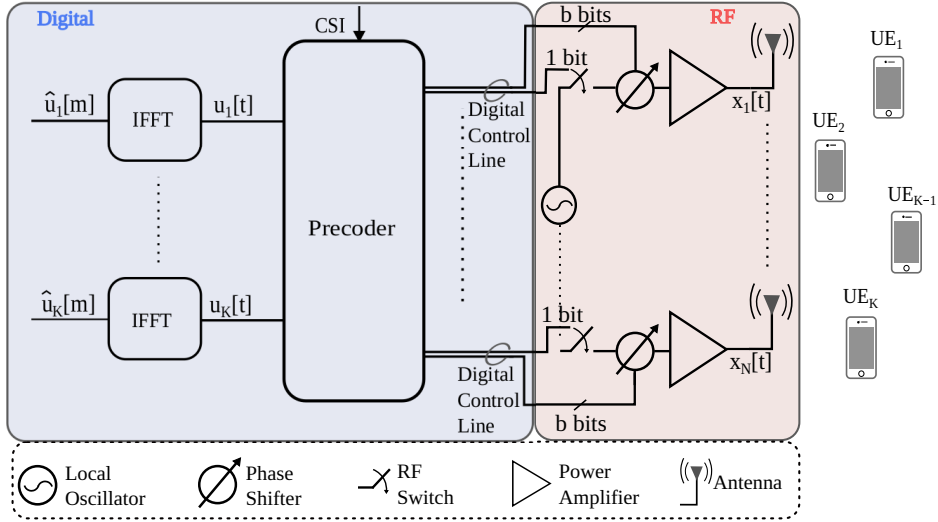}}
	\caption{Multi-user MIMO downlink with a low resolution digitally controlled analog architecture.}
	\label{fig:system}
\end{figure*}

Fig.~\ref{fig:system} shows a \ac{MU-MISO} system with $N$ transmit antennas and $K$ \acp{UE} that each have a single antenna. The base station has one message per \ac{UE} and each antenna has a resolution of 1 bit for the amplitude (on-off switch) and $b$ bits for the phase per antenna. All other hardware components are ideal: linear, infinite bandwidth, no distortions except for \ac{AWGN}. 

\subsection{Baseband Channel Model}
\label{sec:bb_channel_model}
The discrete-time baseband channel is modeled as a finite impulse response filter between each pair of transmit and receive antennas. Let $x_n[t]$ be the symbol of transmit antenna $n$ at time $t$ and let $\vx[t]=(x_1[t]\;\dots\;x_N[t])^T$. Similarly, let $y_k[t]$ be the received symbol of \ac{UE} $k$ at time $t$ and let $\vy[t]=(y_1[t]\;\dots\;y_K[t])^T$. The channel model is
\begin{align}
	\vy[t] =\sum^{L-1}_{\tau=0}\vH[\tau]\vx[t-\tau] + \vz[t]
	\label{eq:FilterChan}
\end{align}
where the noise $\vz[t]=(z_1[t]\;\dots\;z_K[t])^T$ has circularly-symmetric, complex, Gaussian entries that are independent and have variance $\sigma^2$, i.e., we have $\vz \sim \cCN(\zeros,\sigma^2 \vI)$. The $\vH[\tau]$, $\tau=0,\dots,L-1$, are $K \times N$ matrices representing the channel impulse response, i.e., we have
\begin{equation}
\vH[\tau] =
\begin{pmatrix}
h_{11}[\tau] & h_{12}[\tau] & \ldots
& h_{1N}[\tau] \\
h_{21}[\tau] & h_{22}[\tau] & \ldots
& h_{2N}[\tau] \\
\vdots & \vdots & \ddots
& \vdots \\
h_{K1}[\tau] & h_{K2}[\tau] & \ldots
& h_{KN}[\tau]
\end{pmatrix}
\end{equation}
where $h_{kn}[.]$ is the channel impulse response from the $n$-th antenna at the base station to the $k$-th UE. For instance, a Rayleigh fading multi-path channel with a uniform \ac{PDP} has $h_{kn}[\tau] \sim \cCN(0,1/L)$ and these taps are \ac{iid} for all $k,n,\tau$.

The vector $\vx[t]$ is constrained to have entries taken from a discrete and finite alphabet
\begin{equation}
\cX = \{0\} \cup \left\{\sqrt{\frac{P}{N}}\,\mathrm{e}^{\imj 2\pi q/2^{b}} ; q=0,1,2,\dotsc,2^b-1\right\}.
\label{eq:setX}
\end{equation}
The transmit energy clearly satisfies $\norm{\vx[t]}^2\le P$ and we define $\SNR = P/\sigma^2$. The inequality is due to the $0$ symbol that permits antenna selection. Antenna selection was also used in \cite{Zhang16} to enforce sparsity. Our intent is rather to allow antennas not to be used if they do not improve performance. 

\subsection{OFDM Signaling}
\label{sec:ofdm}
Fig.~\ref{fig:system} shows how OFDM can be combined with the precoder. Let $T=T_F+T_c$ be the \ac{OFDM} blocklength with $T_F$ symbols for the DFT and $T_c$ symbols for the cyclic prefix. We assume that $T_F\ge L$ and $T_c\ge L-1$. For simplicity, all $T_F$ subcarriers carry data and we do not include the cyclic prefix overhead in our rate calculations below, i.e., the rates in \ac{bpcu} are computed by normalizing by $T_F$.

Consider the frequency-domain modulation alphabet $\cUhat$ that has a finite number of elements, e.g., QPSK has $\cUhat=\{\hat{u}:\hat{u}=(\pm 1 \pm \imj)/\sqrt{2}\}$. Messages are mapped to the frequency-domain vectors $\hat{\vu}[m] = (\hat{u}_1[m], \dotsc, \hat{u}_K[m])^\tT$ for subcarriers $m=0,\dots,T_F-1$ that are
converted to time-domain vectors $\vu[t]$ by IDFTs
\begin{equation}
	u_k[t] = \frac{1}{T_F}\sum_{m=0}^{T_F-1}\hat{u}_k[m] e^{\imj 2\pi mt/T_F}
\end{equation}
for times $t=0,\dots,T_F-1$ and \acp{UE} $k=1,\dots,K$. For the simulations below, we generated the $\hat u_k[m]$ uniformly from finite constellations such as 16-QAM or 64-QAM. We assume that $\text{E}[\hat u_k[m]]=0$ for all $k$ and $m$.
Each \ac{UE} $k$ uses a DFT to convert its time-domain symbols $y_k[t]$ to the frequency-domain symbols
\begin{equation}
	\hat{y}_k[m] = \sum_{t=0}^{T_F-1} y_k[t] e^{-\imj 2\pi mt/T_F}.
\end{equation}
\subsection{Linear MMSE Precoding}
\label{sec:lmmse}
To describe the linear \ac{MMSE} precoder, consider the channel from base station antenna $n$ to \ac{UE} $k$:
\begin{align}
    \vh_{kn}=(h_{kn}[0], \dotsc,h_{kn}[L-1],\underbrace{0,\dotsc,0}_{\text{$(T_F-L)$ zeros}})^\tT
\end{align}
and denote its DFT
as $\hat{\vh}_{kn}=(\hat{h}_{kn}[0], \dotsc,\hat{h}_{kn}[T_F-1])^\tT$. The channel of subcarrier $m$ is the $K\times N$ matrix $\hat{\vH}[m]$ with entries $\hat{h}_{kn}[m]$ for $k=1,\dots,K$, $n=1,\dots,N$. The linear \ac{MMSE} precoder (or Wiener filter) for subcarrier $m$ is
\begin{align}
    P[m] \hat{\vH}[m]^\dag \left( P[m] \hat{\vH}[m] \hat{\vH}[m]^\dag + \sigma^2\vI \right)^{-1}
    \label{eq:lmmse-matrix}
\end{align}
where $P[m]=\text{E}[|\hat{u}_k[m]|^2]$ is the same for all $k$, $\hat{\vH}[m]^\dag$ is the Hermitian of $\hat{\vH}[m]$, and $\vI$ is the $K\times K$ identity matrix. The precoder multiplies $\hat{\vu}[m]$ by \eqref{eq:lmmse-matrix} for all subcarriers $m$, and performs $N$ IDFTs to compute the resulting $\vx[0],\dots,\vx[T_F-1]$. We remark that \ac{ZF} precoding is the same as \eqref{eq:lmmse-matrix} but with $\sigma^2=0$, where $\hat{\vH}[m] \hat{\vH}[m]^\dag$ is usually invertible if $N$ is much larger than $K$.

\section{Quantized Precoding}
\label{sec:freq_sel_chan}
We wish to ensure compatibility with respect to LP-ZF. In other words, each receiver $k$ should ideally see signals $u_k[t]$, $t=0,\dots,T-1$, that were generated from the frequency-domain signals $\hat u_k[m]$, $m=0,\dots,T_F-1$. Let $\vu[t]=(u_1[t]\;\dots\;u_K[t])^T$ and define the time-domain mean square error (MSE) cost function
\begin{align}
	G(\vx[0], \dotsc,\vx[T-1],\alpha)
	& = \sum^{T-1}_{t=0} \text{E}_{\vz[t]}\left[ \norm {\vu[t] -\alpha \vy[t]}^2 \right]\nonumber \\
	 =\sum^{T-1}_{t=0} \norm {\vu[t] -\alpha \sum^{L-1}_{\tau=0}\vH[\tau]\vx[t-\tau] }^2 & + \alpha^2 TK\sigma^2
	\label{eq:cost-function}
\end{align}
where $\text{E}_{\vz[t]}[\cdot]$ denotes expectation with respect to the noise $\vz[t]$.
The optimization problem is as follows:
\begin{equation}
	\begin{aligned}
		& \underset{\vx[0],\dots,\vx[T-1],\,\alpha}{\text{min}}
		& & G(\vx[0], \dotsc,\vx[T-1],\alpha)\\
		& \text{s.t.} & &  \vx[t] \in \cX^N, \; t={0,\dotsc,T-1} \\
		& & &  \alpha >0 .
	\end{aligned}
	\label{eq:FREQOPTSEQ}
\end{equation}

The parameter $\alpha$ in \eqref{eq:cost-function} and \eqref{eq:FREQOPTSEQ} can easily be optimized for fixed $\vx[0],\dots,\vx[T-1]$ and the result is (see~\cite[Eq.~(26)]{Wang-etal-WCOM18})
\begin{align}
	\alpha = \frac{\sum^{T-1}_{t=0} \Real\left(\vu[t] ^\tH \sum^{L-1}_{\tau=0}\vH[\tau]\vx[t-\tau]\right)}{\sum^T_{t=0} \norm{\sum^{L-1}_{\tau=0}\vH[\tau]\vx[t-\tau]}^2 + TK\sigma^2}.
	\label{eq:alpha}
\end{align}
For the MAGIQ and \ac{QCM} algorithms below, we use alternating minimization to find the $\vx[0],\dots,\vx[T-1]$ and $\alpha$. For the linear \ac{MMSE} precoder, we label the $\alpha$ in \eqref{eq:alpha} as $\alpha_{\textrm{WF}}$. 

Observe that we use the same $\alpha$ for all $K$ \acp{UE} because all \acp{UE} experience the same shadowing, i.e., all $K$ \acp{UE} see the same average power. For \ac{UE}-dependent shadowing, a more general approach would be to replace $\alpha$ with a diagonal matrix with $K$ parameters $\alpha_k$, $k=1,\dots,K$, and then modify \eqref{eq:cost-function} appropriately. 

\subsection{MAGIQ and \ac{QCM}}
\label{sec:algoFS} 
For multipath channels, the vector $\vx[t]$ influences the channel output at times $t, t+1, \dotsc,t+L-1$. A joint optimization over strings of length $T$ seems difficult because of this influence and because of the finite alphabet constraint for the $x_n[t]$. Instead, MAGIQ splits the optimization into sub-problems with reduced complexity by applying coordinate-wise minimization across the antennas and iterating over the OFDM symbol.

For this purpose, consider the precoding problem for time $t'$ starting at $t'=0$ and ending at $t'=T-1$. Observe that $\vx[t']$ influences at most $L$ summands in \eqref{eq:cost-function}, namely the summands for $t=(t')_T,\dots,(t'+L-1)_T$ where $(t)_T=\min(t,T-1)$. To compute the new cost after updating the symbol $x_n[t']$, one may thus compute sums of the form
\begin{align}
	\sum_{t=(t')_T,\dots,(t'+L-1)_T} \norm {\vu[t] -\alpha \sum^{L-1}_{\tau=0}\vH[\tau]\vx[t-\tau] }^2
	\label{eq:OPTSplit2}
\end{align}
for $t'=0,\dots,T-1$. In both cases, one computes a first and second sum having the old and new $x_n[t']$, respectively. One then takes the difference and adds the result to \eqref{eq:cost-function} to obtain the updated cost.

We remark that the time-domain cost function \eqref{eq:cost-function} is closely related to the frequency-domain cost functions in~\cite{Jacobsson-etal-ICT18,Tsinos-etal-Access20,Tsinos-etal-WC21}. However, the time-domain approach is more versatile as it can include acyclic phenomena such as interference from previous \ac{OFDM} blocks. The time-domain approach is also slightly simpler because updating the symbol $x_n[t']$ in \eqref{eq:cost-function} or \eqref{eq:OPTSplit2} requires taking the norm of at most $L$ vectors of dimension $K$ for each test symbol in $\mathcal{X}$ while the frequency-domain approach in~\cite[Eq. (17)]{Tsinos-etal-Access20} takes the norm of $T_F$ vectors of dimension $K$ for each test symbol. Recall that $T_F\ge L$, and usually $T_F\ge 10L$ to avoid losing too much efficiency with the cyclic prefix that has length $T_c\ge L-1$.

The MAGIQ algorithm is summarized in Algorithm~\ref{alg:magiq_fs}. MAGIQ steps through time in a cyclic fashion for fixed $\alpha$. At each time $t$, it initializes the antenna set $\cS=\{1,\dots,N\}$ and performs a greedy search for the antenna $n$ and symbol $x_n[t]$ that minimize \eqref{eq:cost-function} (one may equivalently consider sums of $L$ norms as in \eqref{eq:OPTSplit2}). The resulting antenna is removed from $\cS$ and a new greedy search is performed to find the antenna in the new $\cS$ and the symbol that minimizes \eqref{eq:cost-function} while the previous symbol assignments are held fixed. This step is repeated until $\cS$ is empty. MAGIQ then moves to the next time and repeats the procedure. To determine $\alpha$, MAGIQ applies alternating minimization with respect to $\alpha$ and the precoder output $\{\vx[t]: t=0,\dots,T-1\}$. For fixed $\vx[.]$ the minimization can be solved in closed form, see \eqref{eq:alpha} and line 22 of Algorithm~\ref{alg:magiq_fs}.

\begin{algorithm}[t]
	\caption{MAGIQ and QCM Precoding}
	\label{alg:magiq_fs}
	\begin{algorithmic}[1]
		\Procedure{Precode}{Algo,\,$\vH[.]$,\,$\vu[.]$}
		\State $\vx^{(0)}[.]=\vx[.]_{init}$
		\State $\alpha^{(0)}=\alpha_{init}$
		\For{$i=1:I$} // iterate over OFDM block
		\For{$t=0:T-1$}
		\State $\cS=\{1,\ldots,N\}$
		\While{$\cS\ne\varnothing$}
		\If{Algo = MAGIQ} 
		\State $(x^{\star}_{n^\star}, n^\star) = \argmin_{\tilde x_n\in\cX, n \in \cS}$
		\State $\;\; G\left(\vx^{(i)}[0],\dotsc,\vx^{(i)}[t-1],\tilde{\vx},\right.$
		\State $\left. \;\; \vx^{(i-1)}[t+1], \dotsc,\vx^{(i-1)}[T-1],\alpha^{(i-1)}\right)$
		\Else \; // Algo = QCM
		\State $n^\star = \min \cS$ // round-robin schedule
		\State $x^{\star}_{n^\star}= \argmin_{\tilde x_{n^\star}\in\cX}$
		\State $\;\; G\left(\vx^{(i)}[0],\dotsc,\vx^{(i)}[t-1],\tilde{\vx},\right.$
		\State $\left. \;\: \vx^{(i-1)}[t+1], \dotsc,\vx^{(i-1)}[T-1],\alpha^{(i-1)}\right)$
		\EndIf
		\State $x_{n^\star}^{(i)}[t]= x_{n^\star}^\star$ // update antenna $n^\star$ at time $t$
		\State $ \cS \leftarrow \cS \setminus \{n^{\star}\}$
		\EndWhile
		\EndFor
		\State $\alpha^{(i)} = \frac{\sum^{T-1}_{t=0} \Real\left(\vu[t] ^\tH \sum^{L-1}_{\tau=0}\vH[\tau]\vx^{(i)}[t-\tau]\right)}{\sum^T_{t=0} \norm{\sum^{L-1}_{\tau=0}\vH[\tau]\vx^{(i)}[t-\tau]}^2 + TK\sigma^2}$
		\EndFor
		\State \textbf{return} $\vx[.]=\vx^{(I)}[.],\alpha=\alpha^{(I)}$ 
		\EndProcedure
	\end{algorithmic}
\end{algorithm}

Simulations show that MAGIQ exhibits good performance and converges quickly~\cite{Nedelcu-etal-WSA18}. However, the greedy selection considerably increases the computational complexity. We thus replace the minimization over $\cS$ in line 9 of Algorithm~\ref{alg:magiq_fs}) with a round-robin schedule or a random permutation. We found that both approaches perform equally well. The new QCM algorithm performs as well as MAGIQ but with a simpler search and a small increase in the number of iterations.

\begin{figure}[t!]
	\centering
	\begingroup
	\tikzset{every picture/.style={scale=0.88}}
	\begin{tikzpicture}
\begin{axis}[%
scale only axis,
separate axis lines,
ylabel={$\alpha_{QCM}$},
xlabel={$\alpha_{WF}$},
grid=both,
legend style={at={(0.84,0.18)},anchor=north},
legend cell align=left,
ymin=0.23,
ymax=0.46,
xmin=0.23,
xmax=0.4,
ytick={0.25,0.30,...,0.5},
xtick={0.25,0.30,...,0.5},
]
\addplot [color=TUMBeamerRed,only marks]
  table[row sep=crcr]{%
0.252006 0.253087\\
0.261026 0.262928\\
0.269777 0.272667\\
0.278176 0.282237\\ 
0.285950 0.290744\\ 
0.293554 0.299514\\ 
0.300433 0.307532\\ 
0.307250 0.315752\\ 
0.313347 0.322907\\ 
0.319098 0.329750\\ 
0.324434 0.336668\\ 
0.329374 0.342903\\ 
0.333963 0.349334\\ 
0.338139 0.355318\\ 
0.341956 0.361171\\ 
0.345443 0.366122\\ 
0.348656 0.371255\\ 
0.351569 0.376221\\ 
0.354193 0.380528\\ 
0.356631 0.385522\\ 
0.358787 0.389760\\ 
0.360731 0.393877\\ 
0.362488 0.397533\\ 
0.364070 0.400861\\ 
0.365573 0.405325\\ 
0.366800 0.407100\\ 
0.367986 0.410688\\ 
0.369031 0.413688\\ 
0.369982 0.416300\\ 
0.370810 0.418449\\ 
0.371561 0.420794\\ 
0.372233 0.422599\\ 
0.372826 0.424677\\ 
0.373363 0.426345\\ 
0.373848 0.427850\\ 
0.374283 0.429057\\ 
0.374664 0.430554\\ 
0.375007 0.431602\\ 
0.375323 0.433201\\ 
0.375595 0.434148\\ 
0.375839 0.434999\\ 
0.376054 0.435583\\ 
0.376246 0.436013\\ 
0.376423 0.436349\\ 
0.376580 0.437734\\ 
0.376715 0.437451\\ 
0.376842 0.438958\\
};
\addplot [color=black,line width=1pt,dashed]
  table[row sep=crcr]{%
0.252006 0.252006\\
0.376842 0.376842\\
};
\node[draw,fill=white] at (0.252006, 0.24) {$\ac{SNR}=0$dB};
\node[draw,fill=white] at (0.376842, 0.45) {$\ac{SNR}=20$dB};

\end{axis}
\end{tikzpicture}%
	\endgroup
	\caption{$\alpha_{\textrm{QCM}}$ vs. $\alpha_{\textrm{WF}}$ for System A of Table~\ref{Table-2} and 64-QAM.}
	\label{fig:alpha}
\end{figure}
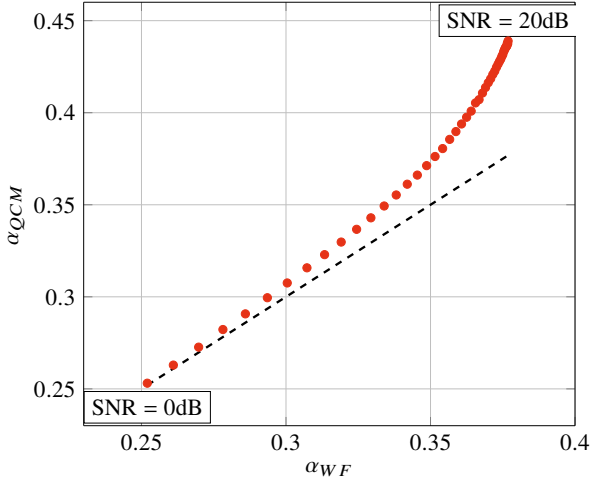

Finally, one might expect that $\alpha$ is close to the $\alpha_{\textrm{WF}}$ of the transmit Wiener filter~\cite{Joh05,bjornson_precoding_2014} since our cost function accounts for the noise power. However, Fig.~\ref{fig:alpha} shows that this is true only at low SNR. The figure plots the average $\alpha$ of the \ac{QCM} algorithm, called $\alpha_{\textrm{QCM}}$, against the computed $\alpha_{\textrm{WF}}$ for simulations with System A in Sec.~\ref{sec:numerical}. Note that $\alpha_{\textrm{QCM}}$ is generally larger than $\alpha_{\textrm{WF}}$.

\section{Performance Metrics}
\label{sec:rcs}
\subsection{Achievable Rates}
\label{sec:rate_computation}

We use \ac{GMI} to compute achievable rates~\cite{kaplan1993information},\cite[Ex.~5.22]{gallager1968} which is a standard tool to compare coded systems. Consider a generic input distribution $P(\vx)$ and a generic channel density $p(\vy|\vx)$ where $\vx=(x_1,\dots,x_S)^T$ and $\vy=(y_1,\dots,y_S)^T$ each have $S$ symbols. A lower bound to the mutual information
\begin{equation}
	\I(\vX;\vY) = \sum_{\vx,\vy} P(\vx) p(\vy|\vx) \log_2\left(\frac{p(\vy|\vx)}{\sum_{\va} P(\va)\, p(\vy|\va)}\right)
	\label{eq:MI}
\end{equation}
is the \ac{GMI}
\begin{align}
	\I_{q,s}(\vX;\vY) = \sum_{\vx,\vy} P(\vx) p(\vy|\vx) \log_2\left(\frac{q(\vy|\vx)^s}{\sum_{\va} P(\va)\, q(\vy|\va)^s}\right) \label{eq:gmi}
\end{align}
where $q(\vy|\vx)$ is any auxiliary density and $s\ge0$. In other words, the choices $q(\vy|\vx)=p(\vy|\vx)$ for all $\vx,\vy$ and $s=1$ maximize the \ac{GMI}. However, the idea is that $p(\vy|\vx)$ may be unknown or difficult to compute and so one chooses a simple $q(\vy|\vx)$. The reason why $p(\vy|\vx)$ is difficult to compute here is because we will measure the \ac{GMI} across the end-to-end channels from the $\hat u_k[m]$ to the $\hat y_k[m]$ and the quantized precoding introduces non-linearities in these channels. The final step in evaluating the \ac{GMI} is maximizing over $s\ge0$. Alternatively, one might wish to simply focus on $s=1$, e.g., see~\cite{arnold_simulation-based_2006}.

We study the \ac{GMI} of two non-coherent systems:\ classic \ac{PAT} and data-aided channel estimation. For both systems, we apply memoryless signaling with the product distribution
\begin{align}
	P(\vx) = 
	\prod_{i=1}^{S_p} 1(x_i=x_{p,i}) \cdot \prod_{i=S_p+1}^{S} P(x_i)
	\label{eq:prod-P}
\end{align}
where the $x_{p,i}$ are pilot symbols, $1(a=b)$ is the indicator function that takes on the value 1 if its argument is true and 0 otherwise, and $P(x)$ is a uniform distribution. Joint data and channel estimation has $S_p=0$ so that we have only the second product in \eqref{eq:prod-P}.
At the receiver we use the auxiliary channel
\begin{align}
	q(\vy|\vx) = \prod_{i=1}^S q_{\vx,\vy}(y_i\,|\,x_i) \label{eq:prod-q}
\end{align}
where the symbol channel $q_{\vx,\vy}(.)$ is a function of $\vx$ and $\vy$. Observe that $q_{\vx,\vy}(.)$ is invariant for $S$ symbols and the channel can be considered to have memory since every symbol $x_{\ell}$ or $y_{\ell}$, $\ell=1,\dots,S$, influences the channel for all ``times'' $i=1,\dots,S$. The \ac{GMI} rate \eqref{eq:gmi} simplifies to
\begin{align}
	\sum_{\vx,\vy} P(\vx) p(\vy|\vx) \sum_{i=S_p+1}^S \log_2\left(\frac{q_{\vx,\vy}(y_i\,|\,x_i)^s}{\sum_{a} P(a)\, q_{\vx,\vy}(y_i\,|\,a)^s}\right). \label{eq:vgmi2}
\end{align}

One may approximate \eqref{eq:vgmi2} by applying the law of large numbers for stationary signals and channels. The idea is to independently generate $B$ pairs of vectors
\begin{align*}
	& \vx^{(b)}=(x_1^{(b)},\dots,x_S^{(b)})^T \\
	& \vy^{(b)}=(y_1^{(b)},\dots,y_S^{(b)})^T
\end{align*}
for $b=1,\dots,B$, and then the following average rate will approach $\I_{q,s}(\vX;\vY)/S$ \ac{bpcu} as $B$ grows:
\begin{align}
	R_\ta = \frac{1}{B} \sum_{b=1}^B R_\ta^{(b)}
	\label{eq:compGMI0}
\end{align}
where
\begin{align}
	R_\ta^{(b)} = \frac{1}{S}
	\sum_{i=S_p+1}^S \log_2\left(\frac{q_{\vx^{(b)},\vy^{(b)}}\left(y_i^{(b)}\,|\,x_i^{(b)}\right)^s}{\sum_{a} P(a)\, q_{\vx^{(b)},\vy^{(b)}}\left(y_i^{(b)}\,|\,a\right)^s}\right). \label{eq:compGMI}
\end{align}

We choose the Gaussian auxiliary density
\begin{equation}
	q_{\vx,\vy}(y|x)
	= \frac{1}{\pi\sigma^2_q} \exp\left(-\frac{\abs{y - h\cdot x}^2}{\sigma_q^2}\right)
	\label{eq:prod-q-symbol}
\end{equation}
where for \ac{PAT} the receiver computes joint \ac{ML} estimates with sums of $S_p$ terms:
\begin{align}
	\begin{split}
		& h = \frac{\sum_{i=1}^{S_p} y_i\cdot x_i^*}{\sum_{i=1}^{S_p} \abs{x_i^2}} \\
		& \sigma_q^2 = \frac{1}{S_p}\sum_{i=1}^{S_p} \abs{y_i - h\cdot x_i}^2 .
	\end{split}
	\label{eq:compParams}
\end{align}
For the data-aided detector we replace $S_p$ with $S$ in \eqref{eq:compParams}. Note that for the Gaussian channel \eqref{eq:prod-q-symbol} the parameter $s$ multiplies $1/\sigma_q^2$ in \eqref{eq:vgmi2} or \eqref{eq:compGMI}, and optimizing $s$ turns out to be the same as choosing the best parameter $\sigma_q^2$ when $s=1$.

Summarizing, we use the following steps to evaluate achievable rates. Suppose the coherence time is $S/T_F$ OFDM symbols where $S$ is a multiple of $T_F$. We index the channel symbols by the pairs $(\ell,m)$ where $\ell$ is the OFDM symbol and $m$ is the subcarrier, $1\le \ell\le S/T_F$, $0\le m\le T-1$. We collect the pilot index pairs in the set $\mathcal{S}_p$ that has cardinality $S_p$, and we write the channel inputs and outputs of \ac{UE} $k$ for OFDM symbol $\ell$ and subcarrier $m$ as $\hat u_k[\ell,m]$ and $\hat y_k[\ell,m]$, respectively. 
\begin{enumerate}
	\item Repeat the following steps (2)-(4) $B$ times; index the steps by $b=1,\dots,B$.
	\item Use Monte Carlo simulation to generate the symbols $\hat u_k[\ell,m]$ and $\hat y_k[\ell,m]$ for $k=1,\dots,K$, $\ell=1,\dots,S/T_F$, and $m=0,\dots,T-1$.
	\item Each \ac{UE} estimates its own channel $h_k$ and $\sigma_{q,k}^2$, i.e., the channel estimate \eqref{eq:compParams} of \ac{UE} $k$ is
	\begin{align}
		\begin{split}
			& h_k = \frac{\sum_{(\ell,m)\in\mathcal{S}_p} \hat y_k[\ell,m] \cdot \hat u_k[\ell,m]^*}{\sum_{(\ell,m)\in\mathcal{S}_p} \abs{\hat u_k[\ell,m]}^2} \\
			& \sigma_{q,k}^2 = \frac{1}{S_p}\sum_{(\ell,m)\in\mathcal{S}_p} \abs{\hat y_k[\ell,m] - h_k\cdot \hat u_k[\ell,m]}^2 .
		\end{split}
		\label{eq:compParams2}
	\end{align}
	For the data-aided detector, in \eqref{eq:compParams2} we replace $\mathcal{S}_p$ with the set of all index pairs $(\ell,m)$, and we replace $S_p$ with $S$.
	\item Compute $R_\ta^{(b)}$ in \eqref{eq:compGMI} for each \ac{UE} $k$ by averaging, i.e., the rate for \ac{UE} $k$ is
	\begin{align}
		R_{\ta,k}^{(b)} =
		\frac{1}{S}
		\sum_{(\ell,m)\notin\mathcal{S}_p} \log_2\left(\frac{q_{\hat \vu_k,\hat \vy_k}\left(\hat y_k[\ell,m]\,|\,\hat u_k[\ell,m]\right)^s}{\sum_{a} P(a)\, q_{\hat \vu_k,\hat \vy_k}\left(\hat y_k[\ell,m]\,|\,a\right)^s}\right)
		\label{eq:compRate}
	\end{align}
	where $\hat \vu_k$ and $\hat \vy_k$ are vectors collecting the $\hat u_k[\ell,m]$ and $\hat y_k[\ell,m]$, respectively, for all pairs $(\ell,m)$. For the data-aided detector we set $\mathcal{S}_p=\emptyset$ in \eqref{eq:compRate}.
	\item Compute $R_\ta$ in \eqref{eq:compGMI0}  for each \ac{UE}, i.e., the average rate of \ac{UE} $k$ is $R_{\ta,k}=\frac{1}{B} \sum_{b=1}^B R_{\ta,k}^{(b)}$.
	\item Compute the average \ac{UE} rate $\overline{R}_{\ta}=\frac{1}{K} \sum_{k=1}^K R_{\ta,k}$.
\end{enumerate}
Our simulations showed that optimizing over $s\ge0$ gives $s\approx1$ if the channel parameters are chosen using~\eqref{eq:compParams2}.
\subsection{Discussion}
\label{sec:rate_discussion}
We make a few remarks on the lower bound. First, the receivers do not need to know $\alpha$. Second, the rate $R_\ta$ in \eqref{eq:compGMI0} is achievable if one assumes stationarity and coding and decoding over many OFDM blocks. Third, as $S$ grows the channel estimate of the data-aided detector becomes more accurate and the performance approaches that of a coherent receiver. Related theory for \ac{PAT} and large $S$ is developed in~\cite{Meng-Gao-Hochwald-A21}. However, the \ac{PAT} rate is generally smaller than for a data-aided detector because the \ac{PAT} channel estimate is less accurate and because \ac{PAT} does not use all symbols for data. We remark that blind channel estimation can approach the performance of data-aided receivers for large $S$. Blind channel estimation algorithms can, e.g., be based on high-order statistics and iterative channel estimation and decoding. For polar codes and low-order constellations, one may use the blind algorithms proposed in~\cite{Yuan-COMML21}. We found that the \ac{PAT} rates are very close (within 0.1 \ac{bpcu}) of the pilot-free rates multiplied by the rate loss factor $1-S_p/S$ for pilot fractions as small as $S_p/S=10\%$.

Depending on the system under consideration, we choose one of $T_F=32,256,396$, one of $T=35,270,277,286,410$, one of $S=256,1584$, and $B=200$. For most simulations we have $T_F=S=256$ and estimate the channel based on individual OFDM symbols, see Sec.~\ref{subsec:contributions}. For example, for $T=270$ and a symbol time of 30 nsec (symbol rate 33.3 MHz) the coherence time needs to be at least $(30\text{ nsec})\cdot T = 8.1$~$\mu$sec. Of course, the transmitter needs to know the channel also, e.g., via time-division duplex, which requires the coherence time to be substantially larger. The main point is that channel estimation at the receiver is not a bottleneck when using \ac{ZF} based on channel inversion.
Finally, for the coded simulations we chose $T_F=396$ and $S=4 T_F=1548$ because the LDPC code occupies four OFDM symbols.

\subsection{Algorithmic Complexity}
\label{sec:complexity}

This section studies the algorithmic complexity in terms of the number of multiplications and iterations. The complexity of SQUID is thoroughly discussed in~\cite{Jacobsson-etal-ICT18} and Table~\ref{Table-1} shows the order estimates take from~\cite[Table~I]{Jacobsson-etal-ICT18}. Note the large number of iterations.

\begin{table*}[t]
	\centering
	\small
	\caption{Algorithmic Complexity}
	\label{Table-1}
	\begin{tabular}{|l|c|c|c|}
		\hline
		Algorithm & \multicolumn{1}{c|}{Multiplications per iteration}  & Iterations   & Pre-processing multiplications                \\ \hline \hline
		QLP-ZF & \multicolumn{1}{c|}{$\cO(TK^3+TK^2N)$} & 1    & - \\ \hline
		SQUID & \multicolumn{1}{c|}{$\cO(8KNT+8NT\log T)$} & 20-300    & $2T\cdot(\frac{5}{3}K^3+3K^2N+(6N-\frac{2}{3})K)$ \\ \hline
		MSM  & \multicolumn{1}{c|}{$ \cO(4KNT^2+4KT+2NT)$} & $\approx$ 8400 & $4KNT$ \\ \hline
		MAGIQ \& QCM &   \multicolumn{1}{c|}{ $\cO(KNTL + KNL|\cX|)$} & 4-6 & $KNT +4NT\log T$   \\ \hline
	\end{tabular}
\end{table*}

The complexity of MSM depends on the choice of optimization algorithm and~\cite{Hela_MSM_OFDM} considers a simplex algorithm. Unfortunately, the simplex algorithm requires a large number of iterations to converge because this number is proportional to the number of variables and linear inequalities that grow with the system size ($N,K,T$). An interior point algorithm converges more quickly but has a much higher complexity per iteration.

For MAGIQ and \ac{QCM}, equation \eqref{eq:cost-function} shows that updating $\vx[.]$ requires updating $L$ of the $T$ terms that each require a norm calculation. The resulting terms $\|\vu[t]\|^2$ do not affect the maximization; terms such as $\norm {\alpha \vH\vx}^{2}_{2}$ can be pre-computed and stored with a complexity of $NKL|\cX|$, and then reused as they do not change during the iterations. On the other hand, products of the form $\alpha \vu^\tH \vH\vx$ must be computed for each of the $L$ terms for each antenna update and at each time instance, resulting in a complexity of $\cO(NKLT)$. The initialization requires $KNT$ multiplications and one must transform the solutions to the time domain. We neglect the cost of updating $\alpha$ because the terms needed to compute it are available as a byproduct of the iterative process over the time instances.

\subsection{Sensitivity to Channel Uncertainty at the Transmitter} 
\label{sec:csi} 

In practice, the \ac{CSI} is imperfect due to noise, quantization, calibration errors, etc. We do not attempt to model these effects exactly. Instead, we adopt a standard approach based on \ac{MMSE} estimation and provide the precoder with channel matrices $\tilde{\vH}[\tau]$ that satisfy
\begin{align}
	\vH[\tau]=\sqrt{1-\varepsilon^2}\tilde{\vH}[\tau] + \varepsilon\vZ[\tau]
\end{align}
where $0\le\varepsilon\le1$ and $\vZ[\tau]$ is a $K \times N$ matrix of independent, variance $\sigma_h^2=1/L$, complex, circularly-symmetric Gaussian entries. Note that $\varepsilon = 0$ corresponds to perfect CSI and $\varepsilon = 1$ corresponds to no CSI. The precoder treats $\tilde{\vH}[\tau]$ as the true channel realization for $\tau=0,\dots,L-1$. 

\section{Numerical Results}
\label{sec:numerical}

We evaluate the \acp{GMI} of four systems. The main parameters are listed in Table~\ref{Table-2} and we provide a few more details here.

\begin{itemize}
	\item System A: the DFT has length $T_F=256$ and the channel has either $L=1$ or $L=15$ taps of Rayleigh fading with a uniform \ac{PDP}. The minimum cyclic prefix length for the latter case is $T_c=14$ so the minimum OFDM blocklength is $T=270$.
	\item System B: MSM is applied to PSK. However, the MSM complexity limited the simulations to smaller parameters than for System A. The channel now has $L=4$ taps of Rayleigh fading with a uniform \ac{PDP}. The $T=35$ OFDM symbols include a DFT of length $T_F=32$ and a minimum cyclic prefix length of $T_c=3$.
	\item System C: System C is actually two systems because we compare the performance under Rayleigh fading to the performance with the Winner2 model~\cite{win2} whose number $L$ of channel taps varies randomly.
	For the Winner2 channel, the choice $T_c=30$ suffices to ensure that $T_c\ge L-1$. The Rayleigh fading model has $L=22$ taps with a uniform \ac{PDP}, where $L$ was chosen as the maximum Winner2 channel length that has almost all the channel energy.
	\item System D: similar to System A but for a 5G NR LDPC code  with code rate 8/9 and 64-QAM for an overall rate of 5.33 \ac{bpcu}. The LDPC code uses the BG1 base graph of the 3GPP Specification 38.212 Release 15, including puncturing and shortening as specified in the standard. The code length is 9504 bits or 1584 symbols of 64-QAM; this corresponds to  4 frames of $T_F=396$ symbols.The codewords were transmitted using at least $T=410$ symbols that include a DFT of length $T_F=396$ and a minimum cyclic prefix length of $T_c=14$.
\end{itemize}

\begin{table*}[t!]
	\centering
	\small
	\caption{System Parameters for the Simulations}
	\label{Table-2}
	\begin{tabular}{|c|r|r|r@{\hspace{.5em}}c@{\hspace{.5em}}l|c|c|c|l|}
		\hline
		System & $N$ & $K$ & $T$&$=$&$T_F+T_c$ & $L$ & Constellation & $b$ & Fading Statistics \\ \hline \hline
		A & 128 & 16 & 270&$=$&$256+14$ & 15 & \{16,\,64\}-QAM & 2,\,3 & Flat and Rayleigh \\ 
		& & & & & & & & & uniform \ac{PDP}\\ \hline
		B & 64 & 8 & 35&$=$&$32+3$ &  4 & \{4-32\}-PSK & 2 & Rayleigh uniform \ac{PDP} \\ \hline
		C & 80 & 8 & 277&$=$&$256+21$ & 22 & 16-QAM & 2 & Rayleigh uniform \ac{PDP} \\
		& & & 286&$=$&$256+30$ & varies & & & Winner2 NLOS C2 urban \\
		\hline
		D & 128 & 16 & 410&$=$&$396+14$ & 15 & 64-QAM & 2 & Rayleigh uniform \ac{PDP} \\
		\hline
	\end{tabular}
\end{table*}

The average \acp{GMI} for Systems A-C were computed using $S=256$, $B=200$, and a data-aided detector. The coded results of System D instead have $S=1584$ symbols to fit the block structure determined by the LDPC encoder. For System D we considered both \ac{PAT} and a data-aided detector. For all cases, the \ac{GMI} was computed by averaging over the sub-carriers, i.e., channel coding is assumed to be applied over multiple sub-carriers and OFDM symbols. The MAGIQ and QCM algorithms were both initialized with a time-domain quantized solution of the transmit matched filter (MF).

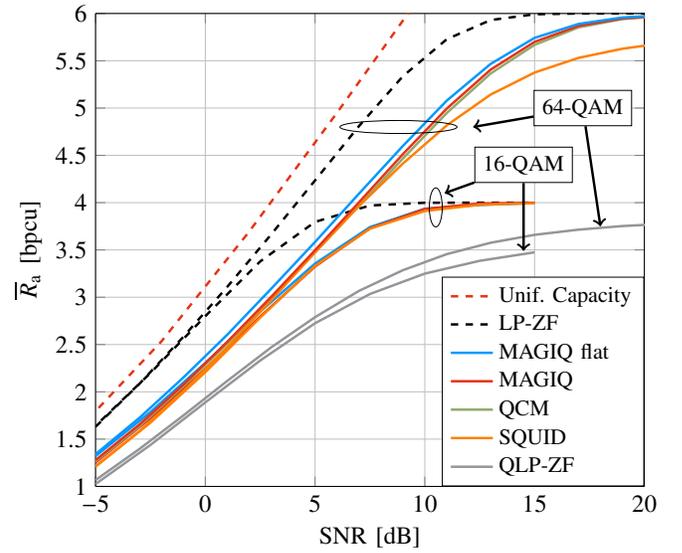
\begin{figure}[t!]
	\centering
	\begingroup
	\tikzset{every picture/.style={scale=0.93}}
	\begin{tikzpicture}
\begin{axis}[%
scale only axis,
separate axis lines,
ylabel={$\overline{R}_\ta$ [\si{\bpcu}]},
xlabel={SNR [\si{dB}]},
grid=both,
legend style={at={(axis cs:20,1)},anchor=south east},
legend cell align=left,
ymin=1,
ymax=6,
xmin=-5,
xmax=20,
ytick={0.5,1,...,6},
]

\addplot [color=TUMBeamerRed,dashed,line width=1.0pt]
  table[row sep=crcr]{%
-10	0.830537007460361\\
-6	1.54955775911384\\
-2	2.53274547220363\\
2	3.69433582568789\\
6	4.95005066967774\\
10	6.24853390612942\\
14	7.56504312867863\\
18	8.88889949547933\\
};
\addlegendentry{\small Unif. Capacity};

\addplot [color=black,dashed,line width=1.0pt]
  table[row sep=crcr]{%
-6.8	1.29154140631597\\
-5	1.64063413962218\\
-2.5	2.19870971950293\\
0	2.79962991380635\\
2.5	3.37859957989772\\
5	3.7963385110003\\
7.5	3.96978999478635\\
10	3.99885272383061\\
12.5	3.99999525151186\\
15    4\\
};
\addlegendentry{\small LP-ZF};

\addplot [color=TUMBeamerBlue,solid,line width=1.0pt]
  table[row sep=crcr]{%
-5	1.31991733813159\\
-2.5	1.78651620714338\\
0	2.30560336310064\\
2.5	2.84830237461647\\
5	3.35629005537125\\
7.5	3.74101169684658\\
10	3.93591768162889\\
12.5	3.975917948176146\\
15	3.996206816288\\
};
\addlegendentry{\small MAGIQ flat};

\addplot [color=TUMBeamerRed,solid,line width=1.0pt]
  table[row sep=crcr]{%
-5	1.25249358008687\\
-2.5	1.72039622697768\\
0	2.24444137297894\\
2.5	2.79603458407619\\
5	3.3242684221211\\
7.5	3.72869213280886\\
10	3.9328921529347\\
12.5	3.99151200942747\\
15  3.99916727782\\
};
\addlegendentry{\small MAGIQ};

\addplot [color=TUMBeamerGreen,solid, line width=1.0pt]
  table[row sep=crcr]{%
-5	1.26471513032228\\
-3	1.63770033115824\\
-1	2.05981013981941\\
1	2.51039095755279\\
3	2.98351296503257\\
5	3.4731130814436\\
7	3.96810558760265\\
9	4.46610855914345\\
11	4.94773664461409\\
13	5.36592404948103\\
15	5.66823436287066\\
17	5.85268590815988\\
19	5.94041496166865\\
21	5.97576675582904\\
23	5.98968774983\\
25	5.99481583502414\\
};
\addlegendentry{\small QCM}

\addplot [color=TUMBeamerOrange ,solid,line width=1.0pt]
  table[row sep=crcr]{%
-5	1.21004351916307\\
-2.5	1.67333879558467\\
0	2.2074948884629\\
2.5	2.78517058320754\\
5	3.33120041034343\\
7.5	3.72428230220072\\
10	3.9115030640865\\
12.5	3.97775792652546\\
15	3.9942048176146\\
};
\addlegendentry{\small SQUID};

\addplot [color=TUMMediumGray,solid,line width=1.0pt]
  table[row sep=crcr]{%
-5	1.0248252862065\\
-2.5	1.43801889128132\\
0	1.88798310482856\\
2.5	2.33077081298764\\
5	2.72500651629123\\
7.5	3.03400230537803\\
10	3.24984547183366\\
12.5	3.38524642990455\\
15      3.47477865543421\\
};
\addlegendentry{\small QLP-ZF};

\addplot [color=black,dashed,line width=1.0pt]
  table[row sep=crcr]{%
-5	1.63052112707042\\
-3	2.08462274539229\\
-1	2.58164150291445\\
1	3.11012128342054\\
3	3.6637791953524\\
5	4.23504006412228\\
7	4.80909388176288\\
9	5.33642491036091\\
11	5.72783695980949\\
13	5.9306485479404\\
15	5.9915669734087\\
17	5.99968174823097\\
19	5.99999741821647\\
21	5.99999999999614\\
23	6\\
25	6\\
};

\addplot [color=TUMMediumGray,solid, line width=1.0pt]
  table[row sep=crcr]{%
-5	1.06699632404432\\
-3	1.39582712768228\\
-1	1.74970599235086\\
1	2.11539675330706\\
3	2.46948400911161\\
5	2.78921809632008\\
7	3.06689948847106\\
9	3.28574479033373\\
11	3.45574012958826\\
13	3.57609919644461\\
15	3.65973079855549\\
17	3.71436001315509\\
19	3.75253135520872\\
21	3.77705806114775\\
23	3.79139249978774\\
25	3.80178065657033\\
};

\addplot [color=TUMBeamerOrange,solid,line width=1.0pt]
  table[row sep=crcr]{%
-5	1.2126852559549\\
-3	1.58444567414094\\
-1	2.00706435229533\\
1	2.47286501511569\\
3	2.96786638623082\\
5	3.47538304522381\\
7	3.96433771625951\\
9	4.41290314108175\\
11	4.81722619703073\\
13	5.14290505992627\\
15	5.37532180441437\\
17	5.53062229203081\\
19	5.62717731405727\\
21	5.69084895906815\\
23	5.72593954781974\\
25	5.74056990982662\\
};

\addplot [color=TUMBeamerRed,solid,line width=1.0pt]
  table[row sep=crcr]{%
-5	1.27752230391153\\
-3	1.65453807861566\\
-1	2.07590998382657\\
1	2.52859175820047\\
3	3.0039402912826\\
5	3.49697244457527\\
7	4.00336718341369\\
9	4.51224021219565\\
11	4.99425985832694\\
13	5.40644189859704\\
15	5.69978278156142\\
17	5.86759754403257\\
19	5.94705842863019\\
21	5.97916145593963\\
23	5.99118921340683\\
25	5.99572062841199\\
};

\addplot [color=TUMBeamerBlue,solid=,line width=1.0pt]
  table[row sep=crcr]{%
-5	1.34049151196965\\
-3	1.72500701191995\\
-1	2.15076028589117\\
1	2.60378116087915\\
3	3.08947043734057\\
5	3.58418898529042\\
7	4.0945015595282\\
9	4.59936495418552\\
11	5.07878877170037\\
13	5.4669582018507\\
15	5.74195660589877\\
17	5.88919618067299\\
19	5.95744406976187\\
21	5.98326262271596\\
23	5.99326741376081\\
23	5.99976741376081\\
};

\node [draw,fill=white] (64qam) at (17.2,5.0) {\small 64-QAM};
\draw [black] (8.8,4.8) ellipse [x radius = 0.90cm, y radius = 0.1cm];
\draw[->,line width=1.0pt] (64qam) -- (12.2,4.8);
\draw[->,line width=1.0pt] (64qam) -- (18,3.75);
\node [draw,fill=white] (16qam) at (14.5,4.4) {\small 16-QAM};
\draw [black] (10.5,3.95) ellipse [x radius = 0.1cm, y radius = 0.3cm];
\draw[->,line width=1.0pt] (16qam) -- (10.9,4.1);
\draw[->,line width=1.0pt] (16qam) -- (14.5,3.49);

\end{axis}
\end{tikzpicture}%
	\endgroup
	\caption{Average \acp{GMI} for System A and $b=2$.}
	\label{fig:MI16QAM_freq_sel}
\end{figure}

Fig.~\ref{fig:MI16QAM_freq_sel} and Fig.~\ref{fig:MI64QAM_ofdm_iter} show the average \acp{GMI} for System A with $b=2$ and $b=3$, respectively. In Fig.~\ref{fig:MI16QAM_freq_sel}, MAGIQ performs 4 iterations for each OFDM symbol while \ac{QCM} performs 6 iterations. Observe that MAGIQ and \ac{QCM} are best at all \acp{SNR} and they are especially good in the interesting regime of high \ac{SNR} and rates. The gap to the rates over flat fading channels ($L=1$) is small. SQUID with $64$-QAM requires 100-300 iterations for $\textrm{SNR}>15$~dB and a modified algorithm with damped updates, otherwise SQUID diverges. In addition, we show the broadcast channel capacity with uniform power allocation and Gaussian signaling as an upper bound for the considered scenario \cite{pviswanath_2003,svishwanath_2003}. Fig.~\ref{fig:MI64QAM_ofdm_iter} shows that \ac{QCM} with 3 iterations operates within $\approx 0.2-0.4$~dB of MAGIQ with 5 iterations when $b=3$ which shows that \ac{QCM} performs almost as well as MAGIQ.

\begin{figure}[t!]
	\centering
	\begingroup
	\tikzset{every picture/.style={scale=0.93}}
	\begin{tikzpicture}
\begin{axis}[%
scale only axis,
separate axis lines,
ylabel={$\overline{R}_\ta$ [\si{\bpcu}]},
xlabel={SNR [\si{dB}]},
grid=both,
legend style={at={(axis cs:19,1.2)},anchor=south east},
legend cell align=left,
ymin=1,
ymax=6,
xmin=-5,
xmax=20,
ytick={0.5,1,...,6},
]

\addplot [color=black,dashed,line width=1.0pt]
  table[row sep=crcr]{%
-5	1.63052112707042\\
-3	2.08462274539229\\
-1	2.58164150291445\\
1	3.11012128342054\\
3	3.6637791953524\\
5	4.23504006412228\\
7	4.80909388176288\\
9	5.33642491036091\\
11	5.72783695980949\\
13	5.9306485479404\\
15	5.9915669734087\\
17	5.99968174823097\\
19	5.99999741821647\\
21	5.99999999999614\\
23	6\\
};
\addlegendentry{\small LP-ZF};

\addplot [color=TUMBeamerRed,solid,line width=1.0pt]
  table[row sep=crcr]{%
-5	1.44931943317034\\
-3	1.85617818374136\\
-1	2.30042470980572\\
1	2.77771245681712\\
3	3.2796545337965\\
5	3.7976344680572\\
7	4.32931408208\\
9	4.86027616999801\\
11	5.33645634487466\\
13	5.69179773240638\\
15	5.89552881150238\\
17	5.97350048353672\\
19	5.99532267162134\\
21	5.99940636188126\\
23	5.99994487981738\\
};
\addlegendentry{\small MAGIQ 5 iter.};

\addplot [color=TUMBeamerGreen,solid,line width=1.0pt]
  table[row sep=crcr]{%
-5	1.40881527531793\\
-3	1.8074543548866\\
-1	2.25478236615195\\
1	2.72869365646781\\
3	3.22650460088591\\
5	3.73853660397567\\
7	4.25944536195129\\
9	4.78206566886071\\
11	5.26853814634982\\
13	5.64956382337574\\
15	5.87234720981462\\
17	5.96836698120667\\
19	5.99467030019757\\
21	5.99942661721324\\
23	5.99996430532943\\
};
\addlegendentry{\small QCM 6 iter.};

\addplot [color=TUMBlue,solid,line width=1.0pt]
  table[row sep=crcr]{%
-5	1.40657076165437\\
-3	1.80225382846831\\
-1	2.24463914950714\\
1	2.71272355665531\\
3	3.20482921915487\\
5	3.71406126280117\\
7	4.23112280890283\\
9	4.74617841502608\\
11	5.22455931861384\\
13	5.6018200098386\\
15	5.83250757305079\\
17	5.94482468742636\\
19	5.98430105725362\\
21	5.99586078681225\\
23	5.99877686089641\\
25	5.99967044299278\\
};
\addlegendentry{\small QCM 3 iter.};

\addplot [color=TUMBeamerOrange,solid,line width=1.0pt]
  table[row sep=crcr]{%
-5	1.40261509085947\\
-3	1.79401933696572\\
-1	2.23059809580697\\
1	2.69324551205837\\
3	3.17890072983166\\
5	3.67896835291398\\
7	4.18116654703733\\
9	4.67117338528826\\
11	5.12326361971103\\
13	5.48128244382605\\
15	5.72022120048492\\
17	5.85941313058877\\
19	5.92745908657876\\
21	5.96003159425068\\
23	5.97501090089457\\
25	5.98243345205842\\
};
\addlegendentry{\small QCM 2 iter.};

\addplot [color=TUMBeamerLightBlue,solid,line width=1.0pt]
  table[row sep=crcr]{%
-5	1.3843098155635\\
-3	1.76291115107681\\
-1	2.18152946468817\\
1	2.61872347696782\\
3	3.06256599107896\\
5	3.4936797103544\\
7	3.89382643660998\\
9	4.24962193758873\\
11	4.54812020568847\\
13	4.77491706991404\\
15	4.93983262994386\\
17	5.05404521987718\\
19	5.13203517294699\\
21	5.18572263454431\\
23	5.21544615148072\\
25	5.23479548950913\\
};
\addlegendentry{\small QCM 1 iter.};

\addplot [color=TUMMediumGray,solid,line width=1.0pt]
table[row sep=crcr]{%
-5	1.17208995095589\\
-3	1.46640325406447\\
-1	1.75317729226923\\
1	2.01283234819777\\
3	2.23081712372428\\
5	2.40220150203208\\
7	2.52740838996765\\
9	2.61752187742156\\
11	2.67878310031597\\
13	2.71806660432438\\
15	2.74561627585735\\
17	2.76013656163312\\
19	2.77272775601994\\
21	2.78108279074036\\
23	2.78412325765772\\
25	2.78572305936359\\
};
\addlegendentry{\small QCM Init.};

\end{axis}
\end{tikzpicture}%
	\endgroup
	\caption{Average \acp{GMI} for System A with 64-QAM and $b=3$.}
	\label{fig:MI64QAM_ofdm_iter}
\end{figure}
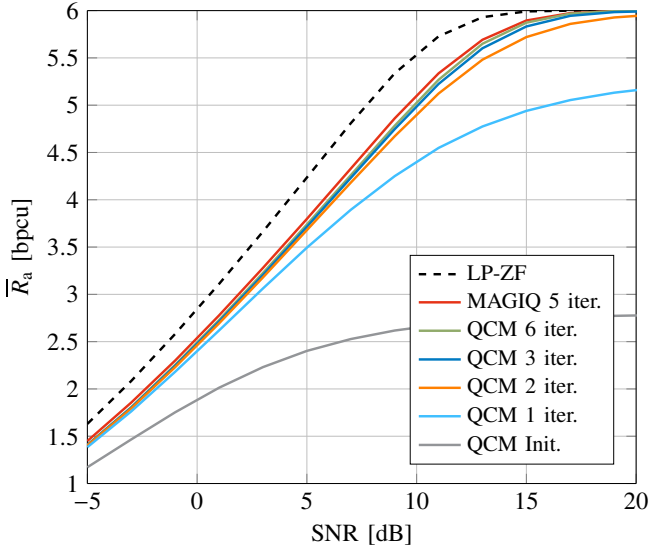

\begin{figure}[t!]
	\centering
	\begingroup
	\tikzset{every picture/.style={scale=0.93}}
	\begin{tikzpicture}
\begin{axis}[%
scale only axis,
separate axis lines,
ylabel={$\overline{R}_\ta$ [\si{\bpcu}]},
xlabel={SNR [\si{dB}]},
grid=both,
legend style={at={(0.92,0.32)},anchor=north},
legend cell align=left,
ymin=1,
ymax=5,
xmin=-8,
xmax=24,
ytick={0.5,1,...,5},
]

\addplot [color=black,dashed,line width=1.0pt]
  table[row sep=crcr]{%
-8	1.12327031057781\\
-7	1.26706034926838\\
-6	1.41384048796088\\
-5	1.5646071175694\\
-4	1.68882034718535\\
-3	1.79806640437184\\
-2	1.88049556278934\\
-1	1.93704236955743\\
0	1.97200624921336\\
1	1.98941931055944\\
2	1.9967207295383\\
3	1.99927434298885\\
4	1.99992311291267\\
5	1.99999774397821\\
6	1.99999992177279\\
};
\addlegendentry{\small LP-ZF};

\addplot [color=TUMBeamerRed,solid,line width=1.0pt]
  table[row sep=crcr]{%
-8	0.819872886464934\\
-7	0.945341681456809\\
-6	1.09240619772731\\
-5	1.22642656492387\\
-4	1.38265582170059\\
-3	1.51523554077612\\
-2	1.63912156737023\\
-1	1.75709003294443\\
0	1.83932054443423\\
1	1.91002079708965\\
2	1.9505549800087\\
3	1.97570257025202\\
4	1.99134965054743\\
5	1.99743076277825\\
6	1.99955124611054\\
};
\addlegendentry{\small QCM};

\addplot [color=TUMBeamerBlue,solid,line width=1.0pt]
  table[row sep=crcr]{%
-8	0.886128949157885\\
-7	1.01951756722445\\
-6	1.1522700091255\\
-5	1.28843211035236\\
-4	1.41597855946946\\
-3	1.55953779480253\\
-2	1.66101903052562\\
-1	1.77253438192396\\
0	1.85676354151942\\
1	1.91623775305576\\
2	1.96251993119391\\
3	1.97980229006996\\
4	1.99185416759465\\
5	1.99545370900621\\
6	1.99960416027194\\\\
};
\addlegendentry{\small MSM};

\addplot [color=TUMBeamerGreen ,solid,line width=1.0pt]
  table[row sep=crcr]{%
-8	0.782569174261793\\
-7	0.917377672451274\\
-6	1.05267287532017\\
-5	1.18755472318375\\
-4	1.34007026835502\\
-3	1.48305318752606\\
-2	1.60691180468053\\
-1	1.73696629567434\\
0	1.82226882513828\\
1	1.89383167347725\\
2	1.94546959534625\\
3	1.97809879043196\\
4	1.99150110634595\\
5	1.99723325282215\\
6	1.99939450829889\\
};
\addlegendentry{\small SQUID};

\addplot [color=black,dashed,line width=1.0pt]
  table[row sep=crcr]{%
-8	1.14384419527479\\
-7	1.30088540694034\\
-6	1.46926927298951\\
-5	1.65670622683378\\
-4	1.82525408125227\\
-3	2.00339315562244\\
-2	2.17705293556142\\
-1	2.34092867550665\\
0	2.49775545508456\\
1	2.63600973730809\\
2	2.75882247993373\\
3	2.85227550521629\\
4	2.91930018501529\\
5	2.96257635852069\\
6	2.98468646703711\\
7	2.99577715792156\\
8	2.9989698461991\\
9	2.99987212267149\\
10	2.99998607430417\\
}; 

\addplot [color=TUMBeamerBlue,solid,line width=1.0pt]
  table[row sep=crcr]{%
-8	0.894208307361201\\
-7	1.04814430275695\\
-6	1.18498330290974\\
-5	1.33902420967587\\
-4	1.48884895224021\\
-3	1.67694075082289\\
-2	1.83160091402757\\
-1	2.01454851879517\\
0	2.17795029665224\\
1	2.31374771959172\\
2	2.46638268761827\\
3	2.59645809078062\\
4	2.71167083989171\\
5	2.81130901796862\\
6	2.87326401926526\\
7	2.91678462102385\\
8	2.9495263216394\\
9	2.97288580082061\\
10	2.98627227060972\\
}; 

\addplot [color=TUMBeamerRed,solid,line width=1.0pt]
  table[row sep=crcr]{%
-8	0.824607503545519\\
-7	0.958055949982445\\
-6	1.11082243652154\\
-5	1.25573571514395\\
-4	1.43348620251682\\
-3	1.59099772199892\\
-2	1.76346849157332\\
-1	1.94259103156634\\
0	2.08509961773876\\
1	2.25417887834048\\
2	2.40735595214918\\
3	2.53623673778577\\
4	2.66239634984043\\
5	2.76705433109058\\
6	2.85381667265876\\
7	2.91568184908514\\
8	2.95124772849298\\
9	2.97770193780374\\
10	2.99076595127876\\
}; 

\addplot [color=TUMBeamerGreen ,solid,line width=1.0pt]
  table[row sep=crcr]{%
-8	0.789232306793746\\
-7	0.926757230904891\\
-6	1.06778447730957\\
-5	1.2148152184933\\
-4	1.37900939048206\\
-3	1.55313879552783\\
-2	1.71781030608897\\
-1	1.89975082736601\\
0	2.05188149443708\\
1	2.21718443080984\\
2	2.37978559403541\\
3	2.52861773212099\\
4	2.6643142913894\\
5	2.76563942560485\\
6	2.84848065761439\\
7	2.90285692947244\\
8	2.94570889111003\\
9	2.9741771822504\\
10	2.98647599665905\\
}; 

\addplot [color=black,dashed,line width=1.0pt]
  table[row sep=crcr]{%
-8	1.14429126707813\\
-6	1.47311083084019\\
-4	1.82544351879934\\
-2	2.18860940260411\\
0	2.52606653439511\\
2	2.86525527202588\\
4	3.20133464533513\\
6	3.51397393794622\\
8	3.77046163377947\\
10	3.92693333388552\\
12	3.98758544380158\\
14	3.99918354011468\\
16	3.99999724823123\\
18	3.99999999990752\\
20	4\\
}; 

\addplot [color=TUMBeamerBlue,solid,line width=1.0pt]
  table[row sep=crcr]{%
-8	0.889159565593449\\
-6	1.19444986738969\\
-4	1.50723486732824\\
-2	1.85059618163781\\
0	2.17706394520104\\
2	2.5258354314322\\
4	2.82920905763283\\
6	3.11139335493189\\
8	3.360627141733\\
10	3.5416211144484\\
12	3.69123013316487\\
14	3.77381294826525\\
16	3.86709783370263\\
18	3.88283655017047\\
20	3.89476932523434\\
}; 

\addplot [color=TUMBeamerRed,solid,line width=1.0pt]
  table[row sep=crcr]{%
-8	0.828130667049706\\
-6	1.0924942137002\\
-4	1.43251926448939\\
-2	1.77298318739888\\
0	2.10610615462222\\
2	2.42232036452321\\
4	2.73755938522222\\
6	3.03386454717935\\
8	3.32858692674281\\
10	3.58341310668597\\
12	3.78783634890113\\
14	3.91393761561428\\
16	3.97508958483334\\
18	3.99425578489485\\
20	3.99859202633102\\
}; 

\addplot [color=TUMBeamerGreen ,solid,line width=1.0pt]
  table[row sep=crcr]{%
-8	0.796992898869172\\
-6	1.08325110757358\\
-4	1.39332212633823\\
-2	1.72506923947401\\
0	2.06684155090054\\
2	2.40071756793372\\
4	2.72043483261992\\
6	3.03884872649415\\
8	3.28732789589083\\
10	3.52689835736415\\
12	3.69681600148214\\
14	3.81103836319091\\
16	3.88460190365384\\
18	3.91468826889179\\
20	3.92632482437374\\
}; 

\addplot [color=black,dashed,line width=1.0pt]
  table[row sep=crcr]{%
-8	1.14429317453071\\
-6	1.47320461423443\\
-4	1.82597178920446\\
-2	2.18826758738305\\
0	2.52635588088013\\
2	2.86535639466527\\
4	3.20478328275738\\
6	3.53605083809471\\
8	3.8717821641755\\
10	4.19964617243224\\
12	4.51738446856616\\
14	4.7727116322387\\
16	4.92699188667974\\
18	4.98757063741058\\
20	4.99914147899919\\
22	4.99999761066489\\
24	4.99999999998226\\
}; 

\addplot [color=TUMBeamerBlue,solid,line width=1.0pt]
  table[row sep=crcr]{%
-8	0.891001363369817\\
-6	1.19211656124716\\
-4	1.50687219298484\\
-2	1.85022553264724\\
0	2.18323829087227\\
2	2.52560156645834\\
4	2.82716710437567\\
6	3.11111237351293\\
8	3.3792749570123\\
10	3.58420132368086\\
12	3.77903417694734\\
14	3.95369593574454\\
16	4.08685957445453\\
18	4.18093437122969\\
20	4.22456923528251\\
22	4.28387172049677\\
24	4.28811822806734\\
}; 

\addplot [color=TUMBeamerRed,solid,line width=1.0pt]
  table[row sep=crcr]{%
-8	0.82788458600224\\
-6	1.10803801156768\\
-4	1.42408423449464\\
-2	1.7682754362458\\
0	2.10075696958533\\
2	2.42238389853736\\
4	2.73488441556381\\
6	3.0369986892288\\
8	3.32280613624817\\
10	3.61904835740113\\
12	3.89524615764516\\
14	4.15688642396566\\
16	4.3983128577907\\
18	4.58537502651174\\
20	4.71728577008147\\
22	4.80677587555095\\
24	4.86116320646469\\
}; 

\addplot [color=TUMBeamerGreen ,solid,line width=1.0pt]
  table[row sep=crcr]{%
-8	0.78593839314295\\
-6	1.07151958440577\\
-4	1.38297805714331\\
-2	1.71668985773898\\
0	2.06508545772248\\
2	2.39830535480432\\
4	2.72282657958383\\
6	3.03696671569681\\
8	3.29381401946999\\
10	3.53040179078438\\
12	3.76264161209589\\
14	3.94377444672125\\
16	4.07751511073057\\
18	4.15525538203844\\
20	4.23089236732528\\
22	4.28515387161218\\
24	4.27982248724841\\
}; 

\node [draw,fill=white] (4psk) at (8.7,2.0) {\small 4-PSK};
\node [draw,fill=white] (8psk) at (12.7,3.0) {\small 8-PSK};
\node [draw,fill=white] (16psk) at (19.0,3.6) {\small 16-PSK};
\node [draw,fill=white] (32psk) at (8.2,4.7) {\small 32-PSK};
\draw [black] (18.5,3.92) ellipse [x radius = 0.1cm, y radius = 0.3cm];
\draw [black,rotate around={165:(16.5,4.4)}] (16.5,4.4) ellipse [x radius = 1.5cm, y radius = 0.1cm];

\end{axis}
\end{tikzpicture}%
	\endgroup
	\caption{Average \acp{GMI} for System B.}
	\label{fig:MSM_freq_sel}
\end{figure}
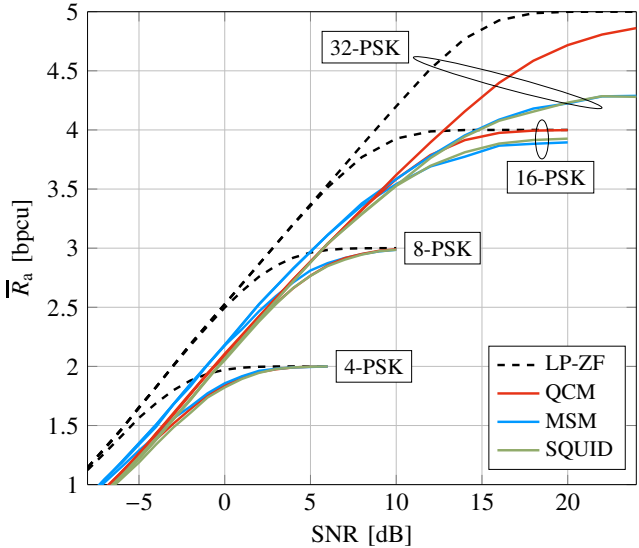

Fig.~\ref{fig:MSM_freq_sel} compares achievable rates of \ac{QCM}, SQUID and MSM  for a smaller system studied in \cite{Hela_MSM_OFDM}. We use PSK because the MSM algorithm was designed for PSK. The figure shows that MSM outperforms SQUID and \ac{QCM} at low to intermediate \ac{SNR} and rates, but \ac{QCM} is best at high \ac{SNR} and rates. This suggests that modifying the cost function \eqref{eq:cost-function} to include a safety margin will increase the \ac{QCM} rate at low to intermediate \ac{SNR}, and similarly modifying the MSM optimization to more closely resemble \ac{QCM} will increase the MSM rate at high \ac{SNR}. We tried to simulate MSM for System A but the algorithm ran into memory limitations (we used 2 AMD EPYC 7282 16-Core processors, 125GByte of system memory, and Matlab with both dual-simplex and interior-point solvers). 

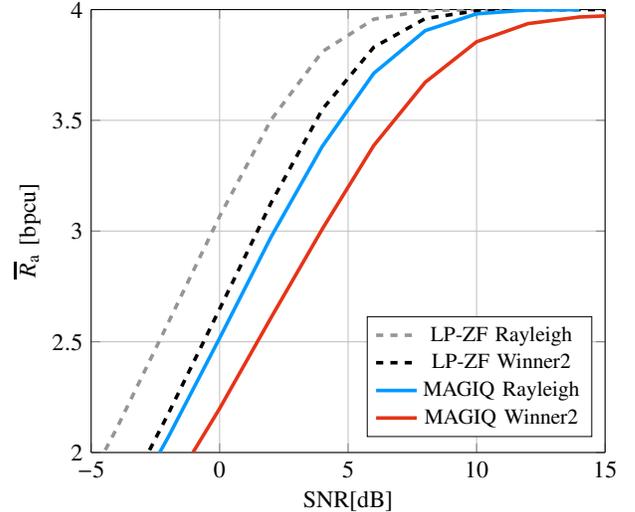
\begin{figure}[t!]
	\centering
	\begingroup
	\tikzset{every picture/.style={scale=0.9}}
	\begin{tikzpicture}
\begin{axis}[
scale only axis,
separate axis lines,
ylabel={$\overline{R}_\ta$ [bpcu]},
xlabel={SNR[dB]},
grid=both,
ymin=2,
ymax=4,
xmin=-5,
xmax=15,
xtick={-5,0,5,10,15},
ytick={0.5,1,...,4},
legend style={at={(axis cs:14.5,2.07)},anchor=south east},
]

\addplot [color=TUMMediumGray,dashed,line width=1.5pt]
  table[row sep=crcr]{%
-6	1.66895836653833\\
-4	2.11018660167653\\
-2	2.58577204560284\\
0	3.06664298516958\\
2	3.50205865980908\\
4	3.8107874595793\\
6	3.95678800890501\\
8	3.99576313050309\\
10	3.99984744865165\\
12	3.99997778314087\\
14	4\\
};
\addlegendentry{\small LP-ZF Rayleigh};

\addplot [color=black,dashed,line width=1.5pt]
  table[row sep=crcr]{%
-4	1.72457123946826\\
-2	2.17447719128434\\
0	2.64811431937591\\
2	3.12680636720292\\
4	3.5498885105096\\
6	3.83039165188658\\
8	3.95953041181178\\
10	3.99478615812185\\
12	3.99983165329425\\
14	3.99999770464474\\
16	3.99999999997003\\
};
\addlegendentry{\small LP-ZF Winner2};

\addplot [color=TUMBeamerBlue,solid,line width=1.5pt]
  table[row sep=crcr]{%
-4	1.65214291643034\\
-2	2.06947220427032\\
0	2.51669153081203\\
2	2.97311414720151\\
4	3.38310757949855\\
6	3.71256506197465\\
8	3.90497542094388\\
10	3.98096886293449\\
12	3.99808313994184\\
14	3.99975977500569\\
};
\addlegendentry{\small MAGIQ Rayleigh};

\addplot [color=TUMBeamerRed,solid,line width=1.5pt]
  table[row sep=crcr]{%
-2	1.81606287713068\\
0	2.19876105365311\\
2	2.60867580425214\\
4	3.01014465974929\\
6	3.38654860867827\\
8	3.6717306539789\\
10	3.85473433344369\\
12	3.93668946045344\\
14	3.96655653873383\\
16	3.97652451009091\\
};
\addlegendentry{\small MAGIQ Winner2};

\end{axis}
\end{tikzpicture}%
	\endgroup
	\caption{Average \acp{GMI} for System C.}
	\label{fig:winner2_f}
\end{figure}

Consider next the Winner2 non-line-of-sight (NLOS) C2 urban model~\cite{win2} that is more realistic than Rayleigh fading. The model parameters are as follows.
\begin{itemize} \itemsep 0pt
	\item Base station at the origin $(x,y)=(0,0)$;
	\item 100 drops of 8 \acp{UE} placed on a disk of radius 150\,m centered at $(x,y)=(0,200\,\mathrm{m})$; the locations of the \acp{UE} are \ac{iid} with a uniform distribution on the disc;
	\item 8x10 uniform rectangular antenna array at the base station with half-wavelength dipoles at $\lambda/2$ spacing;
	\item 5 MHz bandwidth at center frequency 2.53\,GHz;
	\item No Doppler shift, shadowing and pathloss.
\end{itemize}
Fig.~\ref{fig:winner2_f} shows the average \acp{GMI} for LP-ZF and MAGIQ. At high \ac{SNR}, there is a slight decrease in the slope of the MAGIQ \ac{GMI} as compared to LP-ZF. This suggests that one might need a larger $N$ or $b$. The performance for the Rayleigh fading model is better than for the Winner2 model but otherwise behaves similarly. 

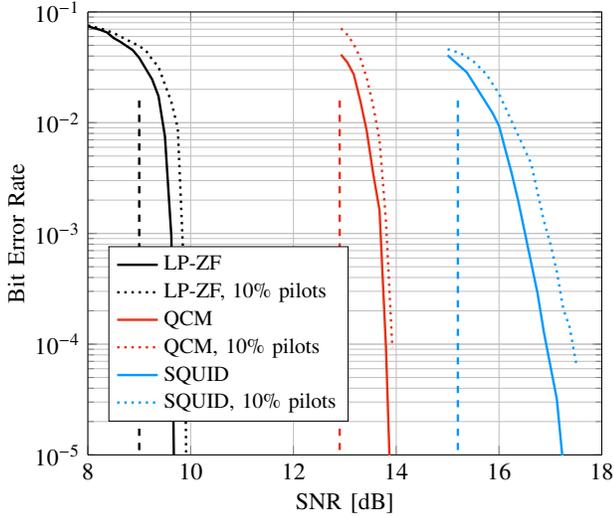
\begin{figure}[t!]
	\centering
	\begingroup
	\tikzset{every picture/.style={scale=0.9}}
	\begin{tikzpicture}
\begin{semilogyaxis}[%
scale only axis,
separate axis lines,
ylabel={Bit Error Rate},
xlabel={SNR [\si{dB}]},
grid=both,
legend pos=south east,
legend cell align=left,
ymin=1E-5,
ymax=0.1,
xmin=8,
xmax=18,
yminorgrids = true,
legend style={at={(axis cs:8.4,2E-5)},anchor=south west}
]

\addplot [color=black,solid,line width=1pt]
table[row sep=crcr]{%
8	    0.07432\\
8.125	0.07145\\
8.25    0.06873\\
8.375   0.06528\\
8.5	    0.05824\\
8.625	0.05404\\
8.75	0.04939\\
8.875	0.04518\\
9	    0.03848\\
9.25	0.02467\\
9.375	0.01736\\
9.5     0.007518\\
9.625	0.000941\\
9.75	0.0000000062\\
};
\draw[dashed,line width=1pt] ({axis cs:9.0,0}|-{rel axis cs:0,0.8}) -- ({axis cs:9.0,0}|-{rel axis cs:0,0});
\addlegendentry{\small LP-ZF};

\addplot [color=black,dotted,line width=1pt]
table[row sep=crcr]{%
8	    0.07561\\
8.125	0.073168\\
8.25    0.07091\\
8.375   0.06777\\
8.5	    0.06463\\
8.625	0.06041\\
8.75	0.05631\\
8.875	0.05274\\
9	    0.04918\\
9.125	0.04510\\
9.25	0.03733\\
9.375	0.03188\\
9.625	0.01538\\
9.75	0.008929\\
9.875   0.0003907\\
10      0.00000000215\\
};
\draw[dashed] ({axis cs:9.0,0}|-{rel axis cs:0,0.8}) -- ({axis cs:9.0,0}|-{rel axis cs:0,0});
\addlegendentry{\small LP-ZF, $10\%$ pilots};

\addplot [color=TUMBeamerRed,solid,line width=1pt]
table[row sep=crcr]{%
12.925	0.04123\\
13.05	0.03502\\
13.175  0.02725\\
13.3    0.01586\\
13.425	0.00851\\
13.55	0.003520\\
13.675 	0.001623\\
13.8	0.0000951\\
13.925  0.0000015\\
};
\draw[TUMBeamerRed,dashed,line width=1pt] ({axis cs:12.9,0}|-{rel axis cs:0,0.8}) -- ({axis cs:12.9,0}|-{rel axis cs:0,0});
\addlegendentry{\small QCM};

\addplot [color=TUMBeamerRed,dotted,line width=1.0pt]
table[row sep=crcr]{%
12.925	0.07109\\
13.05	0.06212\\
13.175  0.05107\\
13.3    0.03865\\
13.425	0.02542\\
13.55	0.01442\\
13.675 	0.00701\\
13.8	0.001304\\
13.925  0.0000925\\
};
\addlegendentry{\small QCM, $10\%$ pilots};

\addplot [color=TUMBeamerBlue,solid,line width=1pt]
table[row sep=crcr]{%
15	    0.0406\\
15.375  0.02846\\
15.875	0.01222\\
16	    0.009375\\
16.25	0.003439\\
16.375	0.001961\\
16.75	0.000288\\
16.875  0.000127\\
17.125  0.000032\\
17.25   0.000008\\
17.375  0.00000072\\
};
\draw[TUMBeamerBlue,dashed,line width=1pt] ({axis cs:15.2,0}|-{rel axis cs:0,0.8}) -- ({axis cs:15.2,0}|-{rel axis cs:0,0});
\addlegendentry{\small SQUID};

\addplot [color=TUMBeamerBlue,dotted,line width=1pt]
table[row sep=crcr]{%
15	    0.04615\\
15.25   0.04241\\
15.5	0.03588\\
15.75	0.02757\\
16	    0.01825\\
16.625	0.004288\\
16.875  0.001286\\
17      0.000808\\
17.125  0.0004526\\
17.25   0.0002090\\
17.375  0.0001424\\
17.5    0.00006766\\
};
\addlegendentry{\small SQUID, $10\%$ pilots};

\end{semilogyaxis}

\end{tikzpicture}%
	\endgroup
	\caption{\acp{BER} for System D and a 5G NR \ac{LDPC} code.}
	\label{fig:LDPC}
\end{figure}

Fig.~\ref{fig:LDPC} shows \acp{BER} for the LDPC code with 64-QAM. Each codeword is interleaved over four OFDM symbols, all 396 subcarriers, and the six bits of each modulation symbol by using bit-interleaved coded modulation (BICM). The interleaver was chosen randomly with a uniform distribution over all permutations of length 9504. The solid curves are for data-aided channel estimation and the dotted curves show the performance of \ac{PAT} when the fraction of pilots is $S_p/S=10\%$. The pilots were placed uniformly at random over the four OFDM symbols and 396 subcarriers. A good blind detector algorithm that performs joint channel and data estimation should have \acp{BER} between the solid and dotted curves.

The dashed curves in Fig.~\ref{fig:LDPC} show the \acp{SNR} required for the different algorithms based on Fig.~\ref{fig:MI16QAM_freq_sel}. In particular, the rate 5.33 bpcu requires \acp{SNR} of 9~dB, 12.9~dB, 15.2~dB for LP-ZF, \ac{QCM} and SQUID, respectively. SQUID is run with 300 iterations and \ac{QCM} is run with 6 iterations. Each UE computes its log-likelihoods based on the parameters~\eqref{eq:compParams} of the auxiliary channel. The \ac{GMI} predicts the coded behaviour of the system within approximately 1~dB of the code waterfall region, except for SQUID where the gap is about 2~dB. The gap seems to be caused mainly by the finite-blocklength of the LDPC code, since the smaller gap of approximately 1~dB is also observed for additive white Gaussian noise (AWGN) channels. The sizes of the gaps are different, and the reason may be that the slopes of the \ac{GMI} at rate 5.33 \ac{bpcu} are different, see Fig.~\ref{fig:MI16QAM_freq_sel}. Observe that LP-ZF exhibits the steepest slope and SQUID the flattest at $R_\ta=5.33$ bpcu; this suggests that SQUID's \ac{SNR} performance is more sensitive to the blocklength.

\begin{figure}[!t]
	\centering
	\tikzset{every picture/.style={scale=0.92}}
	\begin{tikzpicture}[spy using outlines={rectangle, magnification=7,connect spies}]

\begin{axis}[%
scale only axis,
separate axis lines,
xmin=0,
xmax=1,
xlabel={$\epsilon\text{ [estimation error variance]}$},
grid=both,
ymin=0,
ymax=6,
ylabel={$\overline{R}_\ta$ [\si{\bpcu}]},
legend style={draw=white!15!black,fill=white,legend cell align=left},
]

\addplot [color=black,dashed,line width=1.0pt]
  table[row sep=crcr]{%
0	5.841\\
0.1	4.859\\
0.2 4.137\\
0.3 3.562\\
0.4	3.097\\
0.5	2.651\\
0.6 2.214\\
0.7 1.789\\
0.8 1.308\\
0.9 0.7132\\
1   0.0\\
};
\addlegendentry{\small LP-ZF};

\addplot [color=TUMBeamerRed,solid,line width=0.75pt]
  table[row sep=crcr]{%
0	5.17\\
0.1	4.137\\
0.2 3.343\\
0.3 2.781\\
0.4	2.323\\
0.5	1.924\\
0.6 1.556\\
0.7 1.183\\
0.8 0.806\\
0.9 0.426\\
1   0.0\\
};
\addlegendentry{\small QCM};

\addplot [color=TUMBeamerBlue,solid,line width=0.75pt]
  table[row sep=crcr]{%
0	4.816\\
0.1	3.924\\
0.2 3.305\\
0.3 2.776\\
0.4	2.326\\
0.5	1.929\\
0.6 1.569\\
0.7 1.1833\\
0.8 0.7893\\
0.9 0.4105\\
1   0.0\\
};
\addlegendentry{\small SQUID};

\coordinate (spypoint) at (axis cs:0.465,1.65);
\coordinate (spyviewer) at (axis cs:0.4,0.62);
\spy[width=1.8cm,height=1.4cm] on (spypoint) in node [fill=white] at (spyviewer);

\end{axis}
\end{tikzpicture}%
	\caption{Average \acp{GMI} for System A and imperfect CSI at $\textrm{SNR}=12$~dB.}
	\label{fig:cherr}
\end{figure}
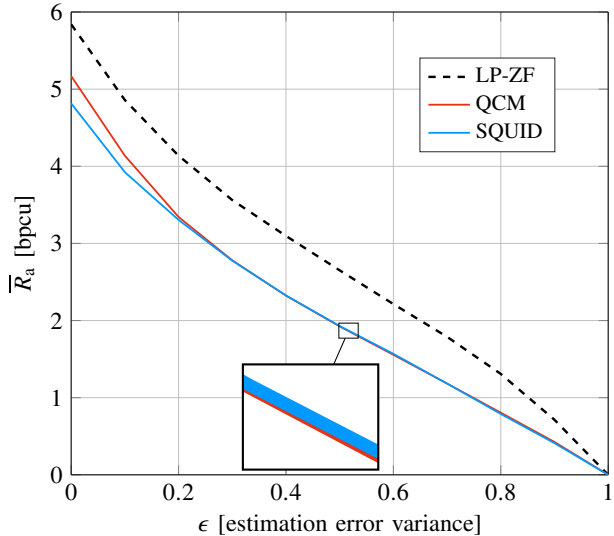

Fig.~\ref{fig:cherr} is for System A and shows how the \ac{GMI} decreases as the \ac{CSI} becomes noisier. The behavior of all systems is qualitatively similar. However, the figure shows that the \ac{QCM} rate is more sensitive to the parameter $\epsilon$ than the SQUID rate when $\epsilon$ is small.

\section{Conclusion}
\label{sec:conclusion}

We studied downlink precoding for MU-MISO channels where the base station uses OFDM and low-resolution DACs. A QCM algorithm was introduced that is based on the MAGIQ algorithm in~\cite{Nedelcu-etal-WSA18} (see also~\cite{Markus17}) and which performs a coordinate-wise optimization in the time-domain. The performance was analyzed by computing the GMI for two auxiliary channel models: one model for pilot-aided channel estimation and a second model for data-aided channel estimation. Simulations for several downlink channels, including a Winner2 NLOS urban scenario, showed that QCM achieves high information rates and is computationally efficient, flexible and robust. The performance of QCM was compared to MAGIQ and other precoding algorithms including SQUID and MSM. The QCM and MAGIQ algorithms achieve the highest information rates with the lowest complexity measured by the number of multiplications. For example, Fig.~\ref{fig:MI64QAM_ofdm_iter} shows that $b=3$ bits of phase modulation operates within 3~dB of LP-ZF. Moreover, \ac{BER} simulations for a 5G NR LDPC code show that \ac{GMI} is a good predictor of the coded performance. Finally, for noisy \ac{CSI} the performance degradation of \ac{QCM} and SQUID is qualitatively similar to the performance degradation of LP-ZF.

\bibliographystyle{IEEEtran}
\bibliography{IEEEabrv,confs-jrnls,./literature}

\end{document}